\documentclass[letterpaper]{article} 
\usepackage{aaai25}
\usepackage{multirow}
\usepackage{subcaption}
\usepackage{amsmath}
\usepackage{booktabs}

\usepackage{times}  
\usepackage{helvet}  
\usepackage{courier}  
\usepackage[hyphens]{url}  
\usepackage{graphicx} 
\usepackage{natbib}  
\usepackage{caption} 
\usepackage{algorithm}
\usepackage{algorithmic}

\usepackage{newfloat}
\usepackage{listings}
\DeclareCaptionStyle{ruled}{labelfont=normalfont,labelsep=colon,strut=off} 
\floatstyle{ruled}
\newfloat{listing}{tb}{lst}{}
\floatname{listing}{Listing}
\title{Counterfactual Inference for Eliminating Sentiment Bias in \\ Recommender Systems}
\author {
    Le Pan\textsuperscript{\rm 1},
    Yuanjiang Cao\textsuperscript{\rm 2},
    Chengkai Huang\textsuperscript{\rm 1}
    Wenjie Zhang\textsuperscript{\rm 1}
    Lina Yao\textsuperscript{\rm 1,2,3}
}
\affiliations {
    \textsuperscript{\rm 1}UNSW\\
    \textsuperscript{\rm 2}Macquarie University\\
    \textsuperscript{\rm 3}CSIRO's Data61\\
    le.pan@unsw.edu.au, caoyuanjiang.learner@gmail.com, chengkai.huang1@unsw.edu.au
    wenjie.zhang@unsw.edu.au
    lina.yao@unsw.edu.au
}

\usepackage{bibentry}

\begin{document}

\maketitle

\begin{abstract}
Recommender Systems (RSs) aim to provide personalized recommendations for users. 
A newly discovered bias, known as sentiment bias, uncovers a common phenomenon within Review-based RSs (RRSs): 
the recommendation accuracy of users or items with negative reviews deteriorates compared with users or items with positive reviews. Critical users and niche items are disadvantaged by such unfair recommendations. 
We study this problem from the perspective of counterfactual inference with two stages. At the model training stage, we build a causal graph and model how sentiment influences the final rating score.
During the inference stage, we decouple the direct and indirect effects to mitigate the impact of sentiment bias and remove the indirect effect using counterfactual inference.
We have conducted extensive experiments, and the results validate that our model can achieve comparable performance on rating prediction for better recommendations and effective mitigation of sentiment bias. To the best of our knowledge, this is the first work to employ counterfactual inference on sentiment bias mitigation in RSs.
\end{abstract}

%

\section{Introduction}
\begin{figure}
\small
    \centering
    \includegraphics[width=1.0\linewidth]{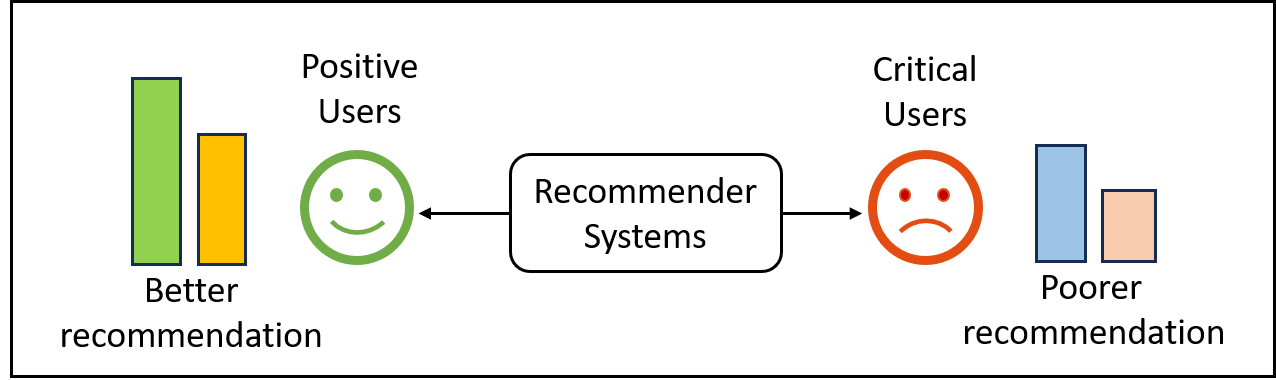}
    \caption{An illustration of sentiment bias.}
    \label{fig:sentiment bias}
\end{figure}

 Recommender Systems (RSs) assist customers or clients in managing the issue of information overload that arises from an overwhelming number of choices \cite{shuaisurvey}. 
Although RSs have a great impact both on industry and academia, they suffer from serious bias issues, which exert negative influences on recommendation performance disparately, both item-side and user-side \cite{chen2023bias,Yao2024}.
Recently, a new bias existing in RSs has been discovered, which is called sentiment bias \cite{lin2021mitigating}. 
It indicates that \textbf{\textit{RSs tend to make more accurate recommendations on users/items having more positive feedback (i.e., positive users/items) than on users/items having more negative feedback (i.e., negative users/items) \cite{lin2021mitigating}}}.
And it also reveals users' emotion and opinions on items are closely concerned with further recommendation performance, as shown in Figure \ref{fig:sentiment bias}.

Sentiment bias is detrimental, which decreases the quality of recommendations to critical users with higher standards and niche items that appeal to a small segment of the user base. On the one hand, critical users are important as they provide in-depth feedback on their unsatisfactory experiences. These negative comments(reviews) benefit the community and help the practitioners of websites or application make improvement if they want to guarantee the user experience and attract more users. Sentiment bias will amplify their negative experiences, resulting in consistently poor recommendations that could deter their ongoing engagement with the platform service. On the other hand, sentiment bias affects niche items by limiting their recommendations, thereby reducing their visibility within the user base \cite{lin2021mitigating} and leading to revenue loss for the platform.

To mitigate sentiment bias, the pioneer work\cite{lin2021mitigating} proposes a heuristic method involving three additional regularization loss terms to the overall optimization objective in RSs. To better reveal the true causal relationships in the recommendation generation process, He et al. \cite{he2022causal} apply causal inference to mitigating sentiment bias.
This method (CISD) builds a causal graph where sentiment is formulated as a Confounder and the Backdoor Adjustment \cite{pearl2009causality} is employed to remove the sentiment bias. 
Since existing RSs datasets like Amazon \cite{Amazon2} do not include explicit sentiment information, incorporating the sentiment variable in the causal graph requires excessive computations with external tools. This increases complexity, potentially impeding real-world applications.
In addition, formulating the sentiment variable as Confounder is unjustified, because sentiment is extracted from user reviews, and user is the cause of review sentiment. Directly removing the effect of sentiment by Backdoor Adjustment might deteriorate the model performance. 

To solve the drawbacks of the current mitigation approaches, 
we leverage the power of counterfactual inference \cite{pearl2009causality}, which provides a novel solution in the realm of RSs debiasing. Counterfactual inference is used to analyze hypothetical scenarios: "what would have happened if certain past conditions or actions had been different". It involves estimating outcomes that were not actually observed. 
Inspired by this, we incorporate counterfactual inference to address sentiment bias, and answer a vital \textit{"What if?"} question: What would the rating score be if RRSs were divested from sentiment bias? Also, counterfactual inference can precisely estimate the specific effect represented as a path in an RSs causal graph by isolating it from other influencing effects. 
For example, if we want to estimate the effect of the user variable in rating prediction, we can construct a counterfactual world where the rating is influenced solely by the user. Given this assumption, we formulate our learning objectives focusing on the direct and indirect effects by creating a neural architecture based on our causal graph. Then we can estimate the effect of sentiment bias, as the effects of user, item, and sentiment are modularized during the training process. In the inference process, we deduct the effect of sentiment from the total predicted rating to mitigate the sentiment bias. Particularly, by employing counterfactual inference, we eliminate the indirect effect, thereby adjusting the ranking score more accurately.
To this end, we leverage causal inference and build a causal graph that does not explicitly require sentiment computation, as sentiment is estimated in the neural architecture of our method. Our approach consists of two stages: (1) We build a causal graph and model how sentiment influences the final rating score during the training stage; (2) During the inference stage, we decouple the direct and indirect effects to mitigate the impact of sentiment bias on the recommendation. 

Our contributions can be summarized as follows:
\begin{itemize}
\item To the best of our knowledge, this is the first work to adopt counterfactual inference on sentiment bias mitigation in RSs;
\item Our work captures the sentiment bias through a causal graph and decreases its impact on inference;
\item Our approach effectively alleviates sentiment bias that is validated by extensive experiments on widely adopted datasets and evaluation metrics.
\end{itemize}

\vspace*{-2.4mm}
\begin{figure*}[ht]
\centering
    \begin{subfigure}[b]{0.2\textwidth}
        \includegraphics[width=0.9\linewidth]{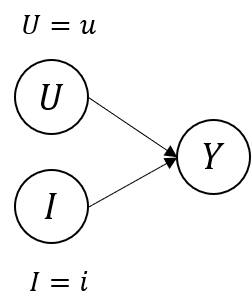}
     \vspace*{-2.4mm}
        \caption{}
        \label{fig:cg-a}
    \end{subfigure}
    \hspace{0.1\textwidth}
    \begin{subfigure}[b]{0.2\textwidth}
        \includegraphics[width=0.9\linewidth]{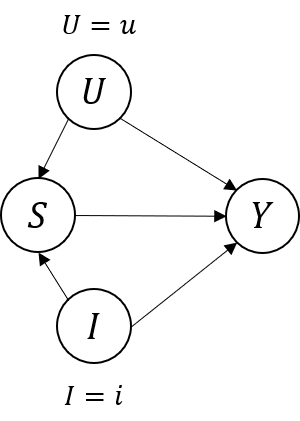}
        \vspace*{-2.4mm}
        \caption{}
        \label{fig:cg-b}
    \end{subfigure}
    \hspace{0.1\textwidth}
    \begin{subfigure}[b]{0.2\textwidth}
        \includegraphics[width=0.9\linewidth]{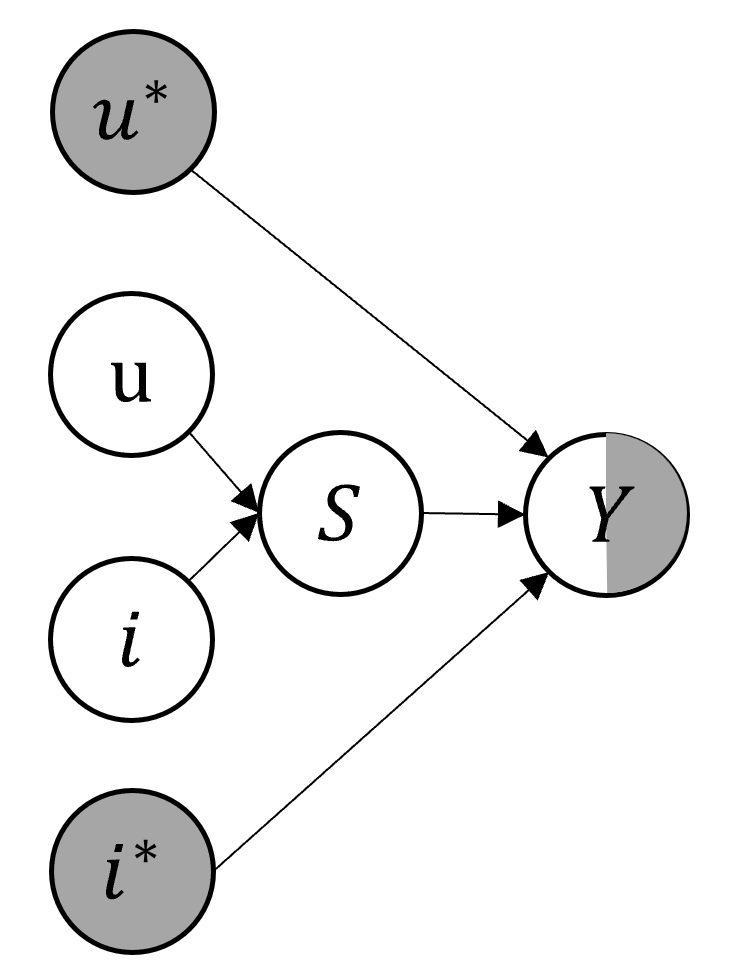}
    \vspace*{-2.4mm}
        \caption{}
        \label{fig:cg-c}
    \end{subfigure}
    \caption{Causal Graph for (a) Traditional user-item matching paradigm. Node $U$ represents the user variable, which refers to the user profile, including review records and interaction history. Node $I$ is the item variable that contains item data, review records and the interaction history. Node $Y$ is the rating variable, which is the output of RRSs. Edge $U \rightarrow Y$ represents the direct effect from user representation to rating. Edge $I \rightarrow Y$ represents the direct effect from item representation to rating. (b) Incorporating sentiment bias; Node $S$ is the sentiment variable that represents the sentiment in the reviews. Edge $U \rightarrow S \rightarrow Y$ and $I \rightarrow S \rightarrow Y$ represent the indirect effect on rating originating from user and item, respectively, with $S$ as the mediator variable. Edge $S \rightarrow Y$ represents the sentiment bias recently proposed by \cite{lin2021mitigating}, which reveals the divergence of recommendation performance between positive user/item and negative user/item. (c) Counterfactual inference. Grey nodes are in the reference state \cite{pearl2009causality}, for example, $u^*$ means $U=u^*$.}
    \label{fig:cg}
\end{figure*}

\section{Related Work}
\subsection{Review-based Recommender Systems (RRSs)}
RRSs adopt text reviews as features or regularizers
to enhance users’ interest prediction where user-item interaction records are inadequate in cold-starting scenarios \cite{sachdeva2020useful}. 
DeepCoNN \cite{zheng2017joint} concatenates all reviews of a given user/item and extracts features from them by a convolution-based network. The rating is predicted based on the interaction between user features and item features. NARRE \cite{chen2018neural} proposes to incorporate an attention mechanism on DeepCoNN structure to decrease the impact of less-useful reviews. MPCN \cite{tay2018multi} proposes a hierarchical attention that considers both review-level and word-level attention to enhance performance. RPRM\cite{wang2021leveraging} explores the usefulness of review properties. They regularize the recommendation loss with a contrastive learning-based loss under the assumption that users would prefer to process information from items of similar usefulness and importance on the review properties.

\subsection{Bias in Recommender Systems}
In this work, we study \textbf{sentiment bias}, which is a type of model bias emerging in the learning process. Limited works on this bias have been proposed so far. Lin et al. \cite{lin2021mitigating} proposed a regularization-based method with three different regularization terms, one for addressing partial item bias, one for the flat distribution of ratings, and the other one for regularizing embeddings.
\cite{xv2022lightweight} proposed LUME to generate a smaller recommendation model based on knowledge distillation and mitigate sentiment bias within Review-based recommender systems (RRSs) simultaneously. \cite{he2022causal} formulated sentiment polarity as a confounder in the causal graph and resolved sentiment bias by causal intervention. This model uses the Backdoor adjustment method to estimate the intervened causal graph during training and then fuses the sentiment term back to the prediction during the inference stage.
Regularization-based alleviation methods typically involve using empirical constraints fail to divest the impact led by sentiment bias. Knowledge distillation requires additional training on a teacher model and a student model, leading to higher time and computation complexity. 


\section{Methodology}
\subsection{Preliminaries}
\label{sec:problem-definition}
Review-based Recommender Systems (RRSs) aim to predict ratings given the input data containing reviews. 
We represent a dataset $\mathcal{D} = \cup^{N}_{k=1} \{(u_k, i_k, \delta^{i_k}_{u_k},y_{u_k,i_k})\}$ consisting of $N$ tuples, where each tuple has a user ID $u$, an item ID $i$, a numerical rating $y_{u,i}$, and a textual review ${\delta}^i_u$ that consists of a sequence of tokens (words). The review and rating $y_{u,i}$ left by a user for an item $i$ reflect this user's attitude towards this item. 
Specifically, given a user $u$ with item $i$ and its textual review ${\delta}^i_u$, RRSs aim to predict the score $y_{u,i}$.

As the core of our model is to leverage counterfactual inference for sentiment bias mitigation, firstly we adopt Directed Acyclic Graphs (DAGs) \cite{shanmugam2015learning} to formulate the causal relations between the variables in RRSs.
For a given DAG, $G=\{V, E\}$, $V$ denotes the node variables set, and $E$ denotes the edges representing the cause-effect relations between variables. An edge points from the cause variable to the effect variable, as shown in Figure \ref{fig:cg}. In Figure (\ref{fig:cg-a}), the traditional user-item causal graph is introduced. This causal graph represents the cause-effect relations for the traditional user-item matching paradigm, where only information from users and items is used to predict the rating, overlooking the sentiment influence embedded in user reviews. Thus, we depict our model in Figure (\ref{fig:cg-b}), where the effect of sentiment bias is incorporated in RRSs. The sentiment node is added to the graph as a mediator variable, constructing two indirect paths towards the ratings, which is more aligned with the real influence in RRSs but has not been discovered by the previous research yet. Figure (\ref{fig:cg-c}) shows the causal graph for counterfactual inference described in the following sections. We can modularize the effect of sentiment in RRSs and better control it for debiasing by Figure (\ref{fig:cg-b}) and Figure (\ref{fig:cg-c}).

\subsection{Counterfactual Inference for Sentiment Debiasing }
\label{ch:counterfactual-inference-sentiment-debias}
As shown in Figure (\ref{fig:cg-b}), user node $U$, item node $I$, and sentiment node $S$ are all the direct causes of rating node $Y$. Thus we obtain the following formulation:
\begin{equation}
  Y_{u,i,s} = Y(U=u, I=i, S=s),  
\end{equation}
where $Y(\cdot)$ means the value function of $Y$, and $u$,$i$,$s$ are the observed values. $S$ is the Mediator Variable, calculated as follows: 
\begin{equation}
  S_{u,i} = S(U=u, I=i),  
\end{equation}
$Y(\cdot)$ and $S(\cdot)$ can be instantiated by neural networks. 
\begin{figure*}[h!]
\small
\centering
        {\includegraphics[width=1.0\linewidth]{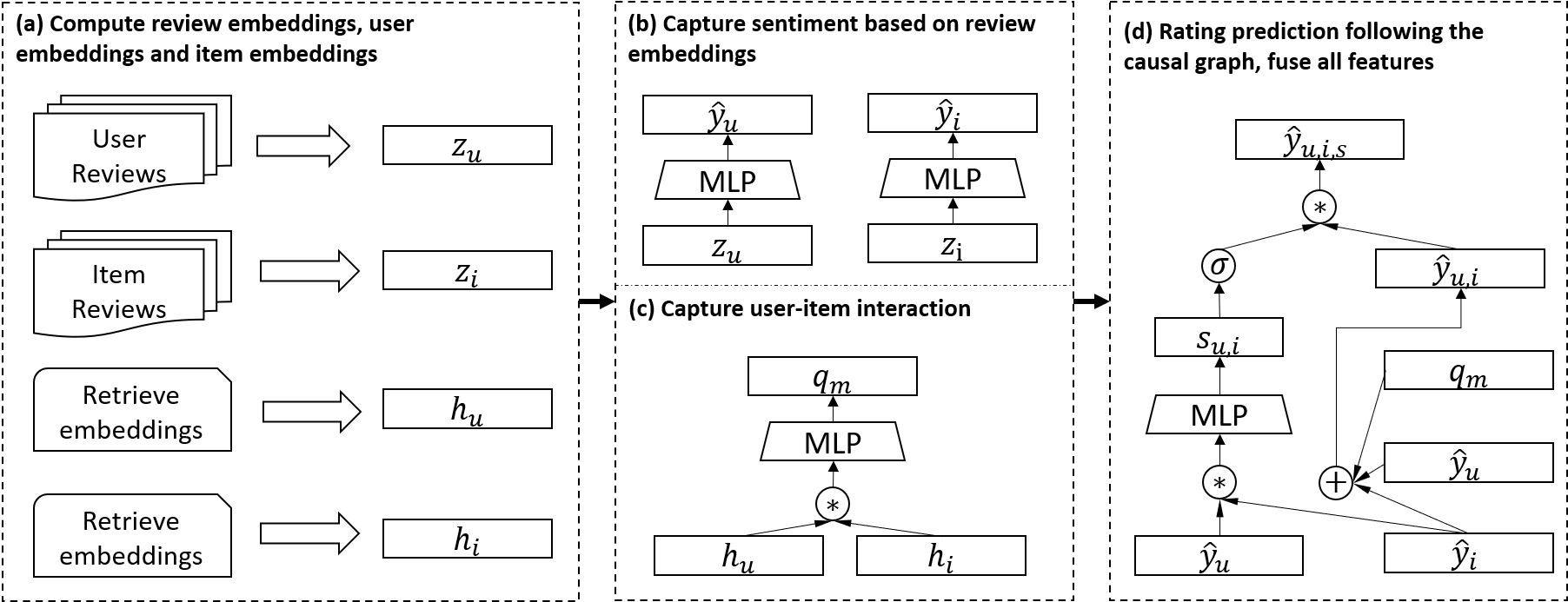}
        }
    \caption{Our proposed RRSs pipeline is based on counterfactual inference. This pipeline consists of (a) embedding computation, (b) capturing sentiment, (c) modelling the user-item interaction, and (d) rating prediction.}
\label{fig:frame}
\end{figure*}
In Figure (\ref{fig:cg-b}), \textit{causal effect} can be computed to estimate and remove sentiment bias. Four paths reflect the causes of $Y$: $U \rightarrow Y$, $I \rightarrow Y$, $U \rightarrow S \rightarrow Y$ and $I \rightarrow S \rightarrow Y$. To alleviate the effect of sentiment bias, we need to remove the effect of $U \rightarrow S \rightarrow Y$ and $I \rightarrow S \rightarrow Y$, and only keep the impact of $U \rightarrow Y$ and $I \rightarrow Y$. Thus, we compute the Natural Direct Effect (NDE), which excludes the indirect effect on $Y$ through the Mediator Variable $S$:
\begin{equation}
    NDE = Y_{u,i,S_{u,i}} - Y_{u^*,i^*,S_{u,i}},
\end{equation}
where $Y_{u,i,S_{u,i}}$ and $Y_{u^*,i^*,S_{u,i}}$ are formulated as:
\begin{equation}
   Y_{u,i,S_{u,i}}=Y(I=i, U=u, S=S(U=u, I=i)) ,
\end{equation}
\begin{equation}
Y_{u^*,i^*,S_{u,i}}=Y(I=i^*, U=u^*, S=S(U=u, I=i)),
\end{equation}
$U=u^*$ and $I=i^*$ indicate the values of $U$ and $I$ are irrelevant to reality and are usually set as null. And when $U$ and $I$ are set as $U=u^*$ and $I=i^*$, NDE represents the differences of $Y$ when $I$ and $U$ change from $i^*$ and $u^*$ to $i$ and $u$, as shown in Figure (\ref{fig:cg-c}).

\subsection{Training}
To implement the counterfactual inference, we build an RRSs model and carry out the training process based on our built causal graph in Figure \ref{fig:cg}. 
Specifically, we add two branches to estimate the causal effect in Path $U \rightarrow S \rightarrow Y$ and $I \rightarrow S \rightarrow Y$ in our model.
$U \rightarrow S \rightarrow Y$ branch takes user reviews as input and computes a value function $Y(U = u)$, which maps text embeddings of reviews to rating through neural networks. This branch captures the user sentiment bias from user reviews by predicting ratings without considering $I$. Similarly, the branch $I \rightarrow S \rightarrow Y$, $Y(I = i)$ predicts ratings without information of $U$, indicating the effect of item sentiment bias. 

We aim to mitigate the effect of sentiment bias based on the above ideas and describe the computation workflow in Figure \ref{fig:frame}. 
Sentiment bias \cite{lin2021mitigating} in RRSs
is correlated with sentiment polarity value, 
which is an assessment derived from the analysis of sentiment within the review text.
Therefore, we capture and mitigate sentiment bias by extracting and exploiting hidden vectors representing sentiment in reviews. 

The user profile in our model consists of two parts: the user reviews $r_u$ and embeddings $h_u$.  A retrieval mechanism maps user IDs and item IDs to continuous dense vector representations, known as user embeddings $h_u$ and item embeddings $h_i$. $h_u$ aims to model the interaction between users and items, together with $h_i$.
 Reviews written by the same user are concatenated to compute the user review embedding $z_u$ using a neural encoder. Firstly, a sequence of word tokens is mapped to word embeddings by word2vec \cite{mikolov2013efficient}. The representation is learned by a simple convolutional block: a convolutional layer with size 5 kernel size, a ReLU layer, a Max-Pooling layer, a fully connected layer, and a dropout layer. The output is piped into a two layer attention module which is composed of a linear layer, a ReLU layer, a dropout layer and finally a linear layer. Attention map is computed and used to attend on salient features. And we carry out the similar computation process for item profiles to obtain $z_i$, as shown in Figure (\ref{fig:frame}a). 
 
Then we capture sentiment based on review embeddings, which show the effect of the two indirect paths on rating as mentioned above. 
The sentiment in user reviews is reflected in the relationship between their embeddings and users' provided ratings, which can be effectively modeled by utilizing review embeddings as input to predict the corresponding ratings.
As shown in Figure (\ref{fig:frame}b), the predicted values $\hat{y}_{u}$ and $\hat{y}_{i}$ are calculated by:
\begin{equation}
    \hat{y}_{u} = f_u(z_u), 
\end{equation}
\begin{equation}
    \hat{y}_{i} = f_i(z_i),
    \label{eq:sentiment-capture} 
\end{equation}
where $f_u(\cdot), f_i(\cdot)$ are neural encoders. $f_u$ maps review embeddings to $\mathcal{Y}$, and $f_i$ maps review embeddings to $\mathcal{Y}$ as well.

Without loss of generality, we employ the classical Neural Collaborative Filtering (NCF) \cite{ncf2017} as backbone to model the user-item interaction. Nevertheless, our sentiment debiasing modeling is model-agnostic and can be integrated with any alternative RSSs model.As shown in Figure (\ref{fig:frame}c), it can be implemented as:
\begin{equation}
    q_m = f_m(h_u \cdot h_i),
\end{equation}
where $f_m(\cdot)$ is a neural operator, and $f_m$ maps user-item interaction information to $\mathcal{Y}$. $q_m$ instantiates the two direct paths in Section \ref{sec:problem-definition}.
Then we fuse all the features based on the causal graph and predict the rating $Y$ by Equation (\ref{eq:sentiment-effect}), where $\hat{y}_{u,i,s}$ denotes the predicted value for rating. The process is shown in Figure (\ref{fig:frame}d): 
\begin{equation}
    \hat{y}_{u,i} = q_m + f_u(z_u) + f_i(z_i), 
\end{equation}
\begin{equation}
    \hat{y}_{u,i,s} = \hat{y}_{u,i} \cdot \sigma(s_{u,i}) \label{eq:sentiment-effect},
\end{equation}
$\hat{y}_{u,i}$ aggregates information from user and item embeddings, user reviews embeddings, and item reviews embeddings. 
 The addition operator implies that the three components are independent, which is consistent with the hypothesis that variables are independent if they are separated in the causal graph \cite{pearl2009causality}. 
Moreover, sentiment $s_{u,i}$ is computed from the multiplication of $z_u$ and $z_i$, manifesting the fuse of sentiment in our causal graph, shown in Equation (\ref{eq:sentiment}):
\begin{equation}
    s_{u,i} = f_s(z_u \cdot z_i) \label{eq:sentiment},
\end{equation}
We use a sigmoid function $\sigma(\cdot)$ to map the combined user-item sentiment $s_{u,i}$ into a value between $(0,1)$, which serves as a control factor for our final rating prediction. Therefore, the computation paths $z_u \rightarrow s_{u,i}$ and $z_i \rightarrow s_{u,i}$ implement the Edge $I \rightarrow S$ and $U \rightarrow S$ in the causal graph. $S \rightarrow Y $ is manifested by $\sigma(s_{u,i})$.

The training objective for RRSs is the Mean Squared Error. 
The classic rating prediction loss $L_{RC}$ is:
\begin{equation}
L_{RC} = \frac{1}{2N} \sum_{u,i} (y_{u,i} - \hat{y}_{u,i,s})^2, \label{eq:rc-mse}
\end{equation}
Similar to MACR \cite{wei2021model}, our model adds two more rating prediction loss functions to help the hidden vectors $z_u$ and $z_i$ extract informative representation from the review data, shown as follows:
\begin{equation}
    L_{U} = \frac{1}{N} \sum_{u,i} (y_{u,i} - \hat{y}_{u})^2, \label{eq:LU}
\end{equation}
\begin{equation}
L_{I} = \frac{1}{N} \sum_{u,i} (y_{u,i} - \hat{y}_{i})^2,  \label{eq:LI}
\end{equation}
The fraction's denominator is $N$ because the two auxiliary loss functions will be fused into the training objective with coefficients. Therefore, compared with Equation (\ref{eq:rc-mse}), the $2$ in the denominators in Equation (\ref{eq:LU}) and Equation (\ref{eq:LI}) are removed.

Finally, we integrate the three loss functions into a multi-task learning objective to train our RRSs model, which follows the causal graph of our proposed model:
\begin{equation}
L = L_{RC} + \alpha_u \cdot L_{U} + \alpha_i \cdot L_{I},
\end{equation}
where $\alpha_u$ and $\alpha_i$ are coefficients that balance the impact of auxiliary loss.

\subsection{Debiased Inference}
As the estimated $\hat{y}_{u,i,s}$ is biased because of the existence of $\sigma(s_{u,i})$, it is necessary to mitigate the sentiment bias in the inference stage. Following the aforementioned counterfactual inference in Section \ref{ch:counterfactual-inference-sentiment-debias}, we implement the inference by:
\begin{equation}
y_{debiased} = \hat{y}_{u,i} \cdot \sigma(s_{u,i}) - \beta \cdot \sigma(s_{u,i}), \label{eq:debias}
\end{equation}
where $\beta \cdot \sigma(s_{u,i})$ corresponds to $Y_{u^*,i^\cdot,S_{u,i}}$ in the NDE formulation, $\beta$ is a reference value of $Y_{u^*,i^*}$. 
$u^*$ and $i^*$ are the reference values of $U$ and $I$. NDE takes effect in this way.

\section{Experiment}
In this section, we perform extensive experiments to verify our model's effectiveness on the widely adopted datasets in RRSs. The experiments are designed to address the following research questions (RQs):

\textbf{RQ1:} Does our proposed 
method improve the model performance on the given datasets?

\textbf{RQ2:} Does our proposed method mitigate the sentiment bias derived from RRSs?

\textbf{RQ3:} How does our proposed method affect the recommendation results?

\textbf{RQ4:} Does our proposed method capture the sentiment bias during computation?

\subsection{Experimental Settings}

\subsubsection{Dataset.} Following previous setting \cite{lin2021mitigating},
we conduct experiments on four different 5-core Amazon datasets \cite{sachdeva2020useful,Amazon1,Amazon2} :
Gourmet Food, Kindle Store, Video Games, Electronics and Yelp\footnote{https://www.yelp.com/dataset} dataset. The user, item and review numbers of these datasets are shown in Table \ref{table:dataset}. 
In the data preprocessing stage, each dataset is randomly split into
training, testing and validation subsets with the proportion of 80\%, 10\% and 10\%, respectively.

\begin{table}[h!]
\begin{center}
\begin{tabular}{cccc}
\toprule
\textbf{Dataset} & \textbf{\#Users} & \textbf{\#Items} & \textbf{\#Reviews} \\ 

\toprule

Gourmet Food & 14,683    & 8,715  & 151,253    \\ 
\midrule
Kindle Store & 68,225    & 61,936 & 982,618    \\ 
\midrule
 Video Games  & 826,769   & 50,212 & 1,324,753  \\ 
\midrule
Electronics  & 192,405   & 63,003 & 1,689,188  \\ 
\midrule
Yelp                & 1,070,074 & 36,490 & 3,766,145  \\ 
\bottomrule
\end{tabular}
\end{center}
\caption{Statistics of the data.}
\label{table:dataset}
\end{table}

\subsubsection{Baselines.} To evaluate our method's effectiveness, we carefully select and compare our proposed model with the following representative non-review-based and RRSs models, 
as well as the two most recent sentiment debiasing methods, Debias\cite{lin2021mitigating} and CISD \cite{he2022causal}. 
Non-review methods include MF \cite{koren2009matrix}, NeuMF \cite{he2017neural}, which are extensively used as baselines in previous works. RRSs models include DeepCoNN \cite{zheng2017joint}, NARRE \cite{chen2018neural}, and MPCN \cite{tay2018multi}.

\subsubsection{Evaluation Metrics.} Following previous works \cite{lin2021mitigating,he2022causal}, we adopt three commonly used evaluation metrics for RRSs debias evaluation: Mean Square Error (MSE), User sentiment Bias (BU), and Item sentiment Bias (BI). The definitions of these metrics are as follows: 
\begin{itemize}
\item MSE: It is selected because most related works, Debias\cite{lin2021mitigating} and CISD \cite{he2022causal}, have both used this same evaluation metric. 
\begin{equation}
    M S E=\frac{1}{N} \sum_{n=1}^N\left(y_n-\hat{y}_n\right)^2,
\end{equation}
where $y_n$ is the $n$-th observed value, $\hat{y}_n$ is the $n$-th predicted value and $N$ is the total number of observations.
\item BU and BI: The user and item sentiment bias for an RRSs model can be defined as:
\begin{equation}
\hspace{-3.7mm}
    B U(R R S) =E\left(R R S, \mathcal{U}^{-}, \mathcal{I}\right)-E\left(R R S, \mathcal{U}^{+}, \mathcal{I}\right), 
\end{equation}
\begin{equation}
\hspace{-3.7mm}
    B I(R R S) =E\left(R R S, \mathcal{U}, \mathcal{I}^{-}\right)- E\left(R R S, \mathcal{U}, \mathcal{I}^{+}\right).
\end{equation}
where $E$ represents the MSE metric. $U$ and $I$ represent the set of users and items, respectively.
The $\mathcal{U}^{+}$ and $\mathcal{U}^{-}$ are positive users and negative users, respectively. They are selected from the top 10\% and bottom 10\% users sorted by the user sentiment polarity scores. Similarly, $\mathcal{I}^{+}$ and $\mathcal{I}^{-}$ are top 10\% and bottom 10\% items sorted by the item sentiment polarity scores.

\end{itemize}

\subsubsection{Implementation Details.}
\begin{figure*}[h]
\centering

\hspace*{-6mm}
\includegraphics[scale = 0.33]{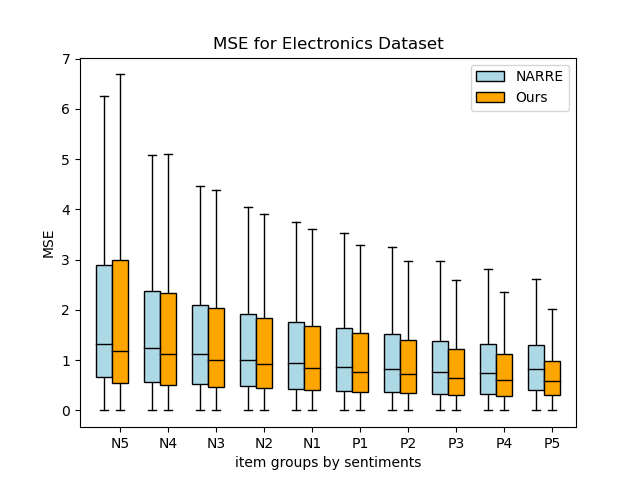}\hspace*{-6mm}
\includegraphics[scale = 0.33]{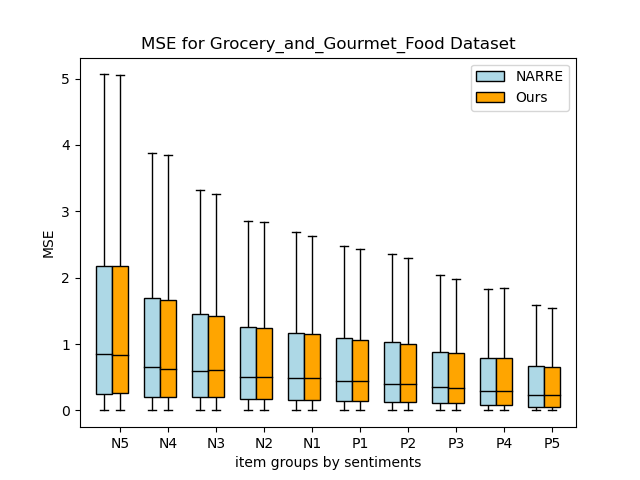}\hspace*{-6mm}
\includegraphics[scale = 0.33]{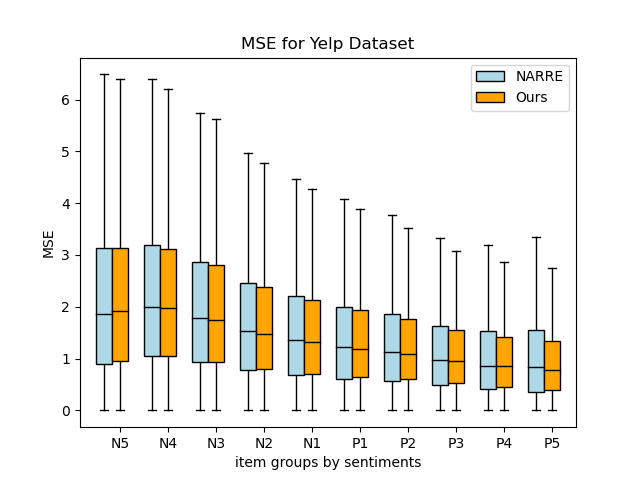}
\vspace*{-2.4mm}
\smallskip

\hspace*{-7mm}
\includegraphics[scale = 0.33]{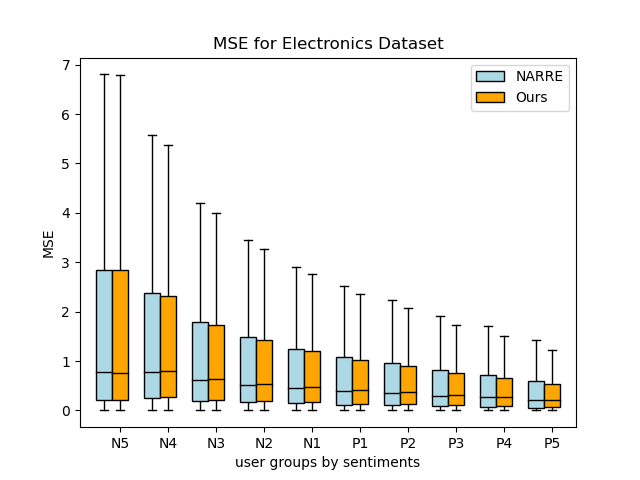} \hspace*{-6.9mm}
\includegraphics[scale = 0.33]{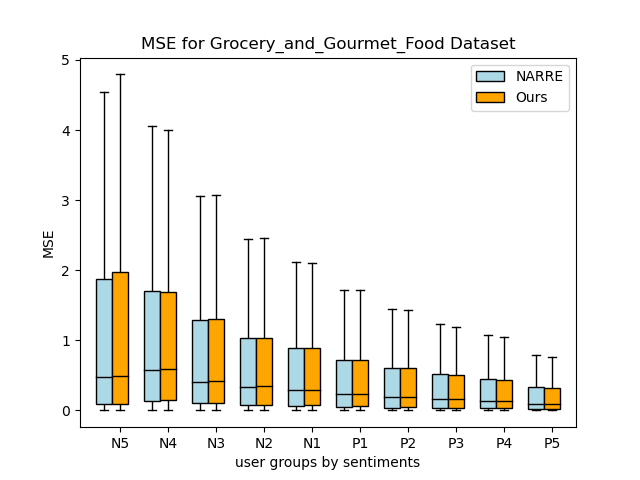} \hspace*{-6.5mm}
\includegraphics[scale = 0.33]{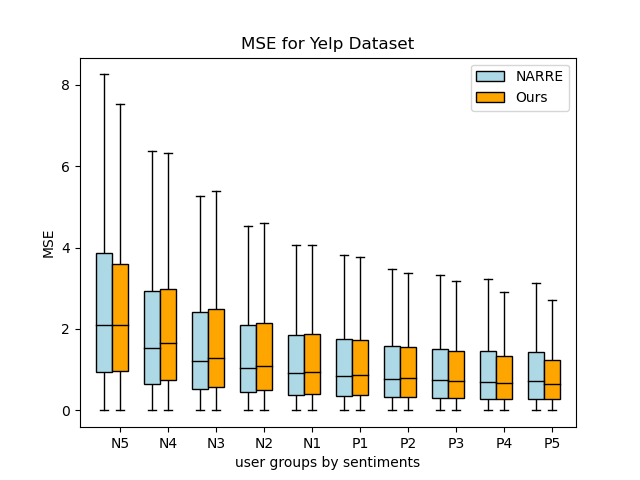} 

\caption{Boxplots on mean MSE comparison before and after debias(RQ3).}
\label{fig:boxplots}
\end{figure*}

In order to maintain the generality of our method, we follow the commonly-used setting in \cite{sachdeva2020useful,lin2021mitigating,he2022causal} for RRSs.
 We optimize the hyper-parameters with the validation set and then use the test datasets to verify our model's efficacy. For fair comparison, we compare the performance of our method with the optimal baseline results.
  All the implementation code, split datasets and hyperparameter setting are provided in supplementary material for reproducibility.






\begin{figure*}[h]
    \centering
    \begin{subfigure}[b]{0.23\textwidth}
        \includegraphics[width=\textwidth]{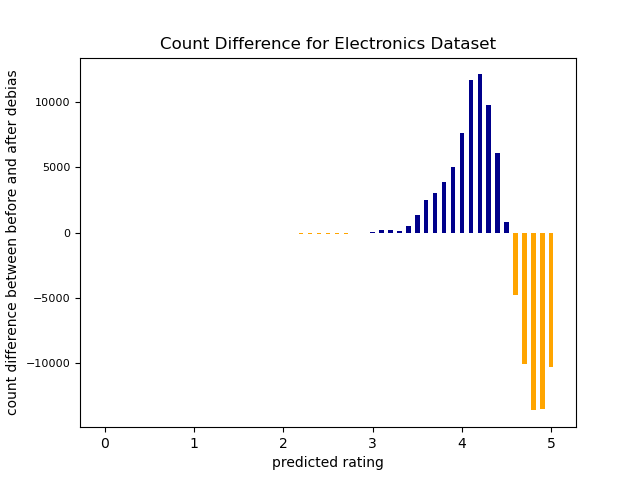}
    \end{subfigure}
    \hfill
    \begin{subfigure}[b]{0.23\textwidth}
        \includegraphics[width=\textwidth]{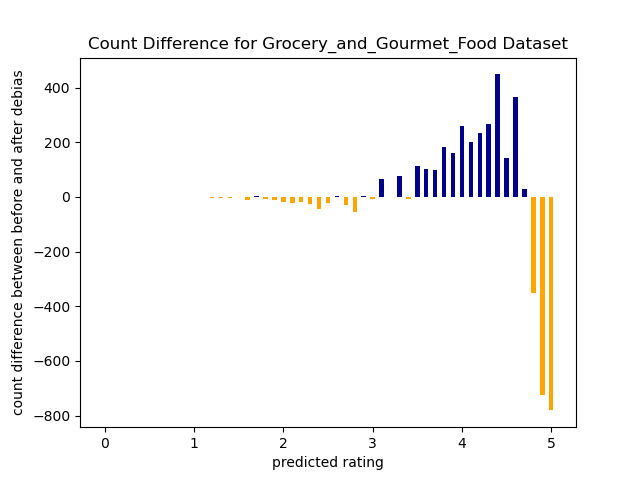}
    \end{subfigure}
    \hfill
    \begin{subfigure}[b]{0.23\textwidth}
        \includegraphics[width=\textwidth]{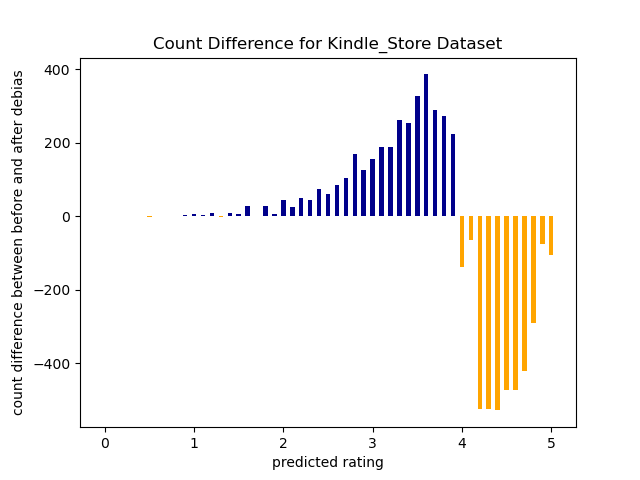}
    \end{subfigure}
    \hfill
    \begin{subfigure}[b]{0.23\textwidth}
        \includegraphics[width=\textwidth]{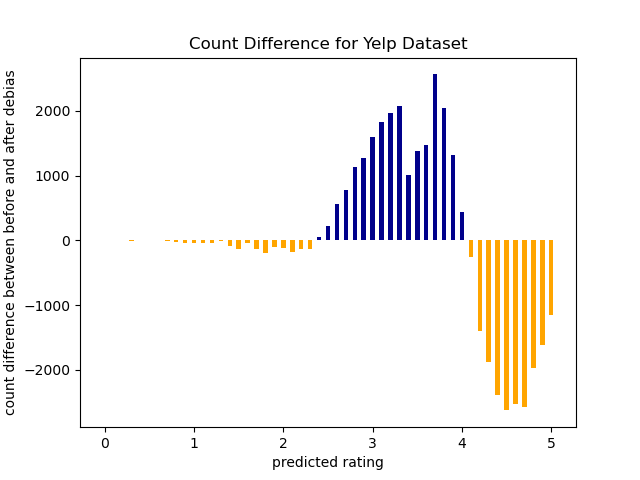}
    \end{subfigure}
    \caption{Difference between before and after debias in rating distribution shift(RQ3).}
    \label{fig:diff-barplot}
\end{figure*}

\begin{figure*}[h!]
\centering
\begin{subfigure}[b]{0.24\textwidth}
    \centering
    \includegraphics[width=\textwidth]{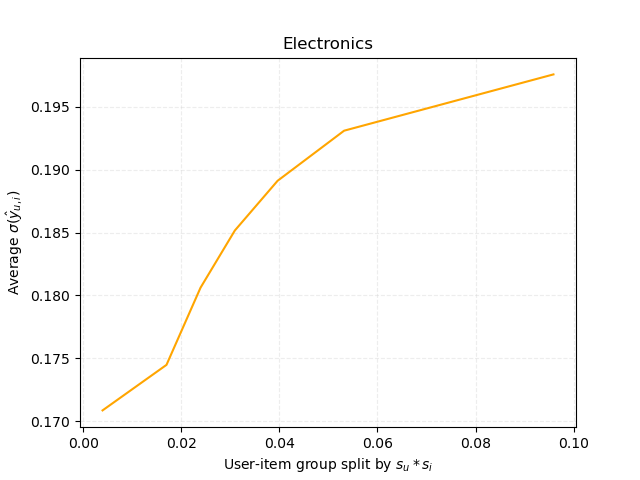}
    \caption{Electronics}\label{fig:senti-sigma-plot-a}
\end{subfigure}
\hfill
\begin{subfigure}[b]{0.24\textwidth}
    \centering
    \includegraphics[width=\textwidth]{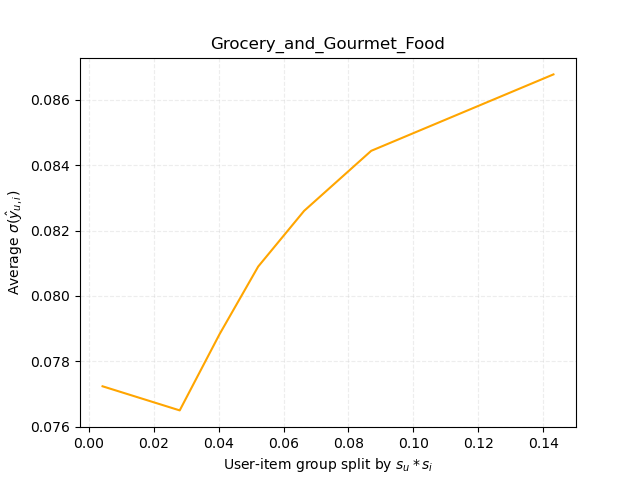}
    \caption{Grocery}\label{fig:senti-sigma-plot-b}
\end{subfigure}
\hfill
\begin{subfigure}[b]{0.24\textwidth}
    \centering
    \includegraphics[width=\textwidth]{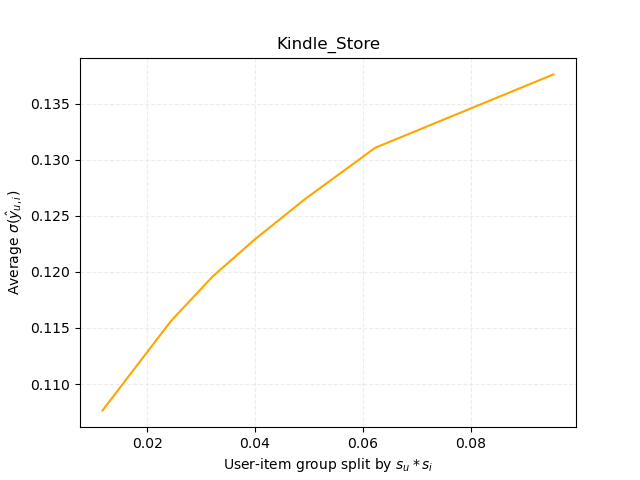}
    \caption{Kindle Store}\label{fig:senti-sigma-plot-c}
\end{subfigure}
\hfill
\begin{subfigure}[b]{0.24\textwidth}
    \centering
    \includegraphics[width=\textwidth]{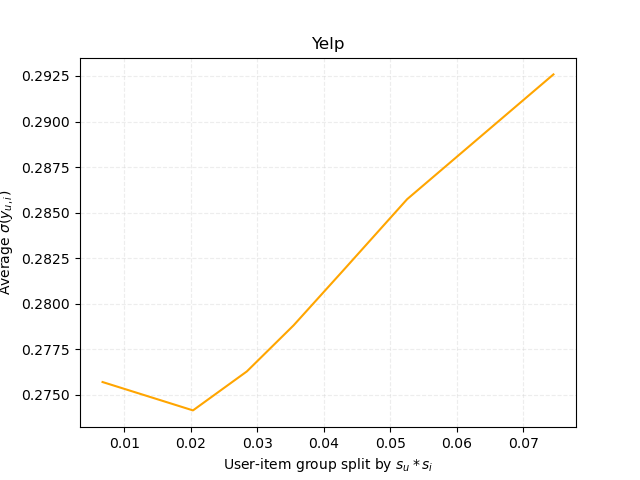}
    \caption{Yelp}\label{fig:senti-sigma-plot-d}
\end{subfigure}

\caption{The relationship between the predicted $\sigma(\hat{y}_{u,i})$ and related sentiment $s_u \cdot s_i$ (RQ4).}
\label{fig:senti-sigma-plot}
\end{figure*}

\subsection{Recommendation Efficacy (RQ1)}

\begin{table}[h!]
\small
\centering
\begin{tabular}{cccccc}
\toprule
\textbf{Model}  & \textbf{Gourmet} & \textbf{Video} & \textbf{Kindle} & \textbf{Elec} & \textbf{Yelp}    \\ 
\midrule
\midrule
MF             & 0.9728  & 1.0962      & 0.7074 & 1.3215      & 1.5207  \\ 
\midrule
NeuMF          & 0.9693  & 1.0965      & 0.713  & 1.3187      & 1.5167  \\ 
\midrule
DeepCo       & 0.9942  & 1.1496      & 0.6962 & 1.2912      & 1.5740   \\ 
\midrule
NARRE          & 0.9669  & 1.0882      & 0.6612 & 1.2588      & 1.5119  \\ 
\midrule
MPCN           & 1.1966  & 1.6608      & 0.9077 & 1.4075      & 1.6718  \\ 
\midrule
De-bias        & 0.9652  & 1.3713      & 0.6244 & 1.2394      & 1.4535  \\ 
\midrule
CISD           & 0.9641  & 1.083       & 0.6104 & \textbf{1.2253}      & 1.4473  \\ 
\midrule
Ours       & \textbf{0.9485}  & \textbf{1.0254}      & \textbf{0.5693} & 1.2357      & \textbf{1.3693} \\ 
\bottomrule
\end{tabular}
\caption{MSE for RSs models on Amazon and Yelp Datasets.}
\label{MSE comparison}
\end{table}

The rating prediction performance of other comparison models and our model are shown in Table \ref{MSE comparison}. The best result on each dataset is presented in bold font. Except for the Electronics dataset, our method achieves state-of-the-art performance on the other four datasets. On the Electronics dataset, our method achieves the second-best performance with a marginal discrepancy with the best performance. On the Kindle and Yelp datasets, our method improves the MSE by a large margin compared with other models, showing the superiority of our proposed method.

\subsection{Sentiment Debiasing Effect (RQ2)}
\label{sec:sentiment-debiasing-effect}

\begin{table}[h!]
\centering
\begin{tabular}{c|c|c|c|c}
\toprule
\textbf{Datasets} & \textbf{Model}   & \textbf{BU} & \textbf{BI} & \textbf{MSE}    \\ 
\midrule
\midrule
\multirow{3}{*}{Gourmet}         & NARRE           & 1.2759  & 0.8067      & 0.9669   \\ 
& De-bias   & 1.2344  & 0.7655      & 0.9652   \\ 
& Ours            & \textbf{1.1843}  & 1.0177      & \textbf{0.9485}   \\ 
\midrule

\multirow{3}{*}{Video Games}         & NARRE      & 2.2774  & 2.1054 & 1.5260   \\ 
& De-bias    & 1.9752  & 1.4594      & 1.4388   \\ 
& Ours            & \textbf{1.1976}  & \textbf{0.6008}      & \textbf{1.026}   \\ 
\midrule

\multirow{3}{*}{Kindle}         & NARRE           & 1.0024  & 0.7044  & 0.6612   \\ 
& De-bias    & 0.9247  & 0.6469    & 0.6244   \\ 
& Ours            & \textbf{0.8584}  & \textbf{0.4693}    & \textbf{0.5694}   \\ 
\midrule

\multirow{3}{*}{Electronics}         & NARRE      & 1.4132  & 1.1890  & 1.2588   \\ 
& De-bias    & 1.3522  & 1.1352      & 1.2394   \\ 
& Ours            & \textbf{1.3507}  & \textbf{1.1093}      & \textbf{1.2357}   \\ 
\midrule

\multirow{3}{*}{Yelp}         & NARRE           & 2.1043   & 1.2952   & 1.5119   \\ 
& De-bias   & 1.7749  & 1.0028      & 1.4535   \\ 
& Ours            & \textbf{1.2867}  & \textbf{0.9450}     & \textbf{1.4003}  \\ 

\bottomrule
\end{tabular}
\caption{BU, BI and MSE for RSS models on Amazon and Yelp Datasets.}
\end{table}

To answer RQ2, we conduct experiments to demonstrate that our strategy effectively mitigates sentiment bias and accounts for improving recommendation performance.
Since CISD\cite{he2022causal} has not released their codes publicly, we contrast our approach with the De-bias method \cite{lin2021mitigating}, which provides detailed comparison results on sentiment bias mitigation. As our implementation is based on NARRE \cite{chen2018neural}, it is also selected for comparison. 
The Debias \cite{lin2021mitigating} method uses three regularization losses to remove the effect of sentiments heuristically.

Still, numbers in bold font represent the best performance. Our method performs the best among almost all the five datasets in terms of BU and BI, except for BI on the Gourmet dataset. On the Video Games dataset,  our method reduces more than 50\% in BU and BI when compared with Debias.
Therefore, it is evident that our approach effectively alleviates sentiment bias.

\subsection{Effect on Recommendation Results (RQ3)}





In the above two research questions (RQ1 and RQ2), we have provided quantitative analysis on recommendation efficacy and sentiment debiasing effect. To further reveal the impact of our mitigation approach on recommendations, we adopt boxplots to compare mean MSE among different groups with different sentiment polarity levels. 
Similar to \cite{lin2021mitigating}, we divide users and items into ten groups based on their sentiment polarity, computed by a lexicon-based analysis tool called  TextBlob\footnote{https://github.com/sloria/TextBlob}. 

As shown in Figure \ref{fig:boxplots}, $P_{i}$ and $N_{i}$ ($i=1,...,5$) denote the groups with positive and negative sentiment.
With the increase of $i$, the sentiment polarity level of each group also increases (more positive or negative). The groups are arranged by sentiment, and we compare our methods with the baseline on three datasets. The graphs demonstrate our method decreases the mean MSE in the majority groups on all datasets, both for user and item groups. The variance of most boxes also decreases, which manifests in the shorter whiskers in the boxplots.
In addition, we illustrate the differences between the distribution of ratings before and after our debias approach in Figure \ref{fig:diff-barplot}. We use the ratings predicted by NARRE (similar to Section \ref{sec:sentiment-debiasing-effect}) as the results before debias, and the ratings from our proposed method as the results after debias.
We compute and visualize the count subtraction difference between the above two results.
As shown in Figure \ref{fig:diff-barplot}, the vertical axis is the count difference, and the horizontal axis is the predicted rating, which are float numbers in the range of $[0,5]$.
 We draw the negative difference in orange and the positive difference in blue.
 In sentiment analysis literature, the positive polarity is defined as 
we can observe a notable positive difference when the predicted ratings range from $(2,4)$, where the sentiment polarity is lower.
 Furthermore, the count difference in ratings below $2$ is smaller than those above $4$. This means our debias model has a larger effect on positive sentiment polarity, which decreases the effect of sentiment bias. 
 
 Therefore, according to Figure \ref{fig:boxplots} and Figure \ref{fig:diff-barplot},  it is reasonable to conclude that our proposed model improves the overall MSE by decreasing average MSE in most groups with different sentiment polarity and the model mitigates the sentiment bias by imposing a different magnitude change in ratings above $4$ and ratings below $2$.

 \subsection{Relationship between Sentiment and Predicted Values (RQ4)}

Unlike CISD \cite{he2022causal}, our method does not use sentiment analysis tools to generate sentiment polarity as extra information to facilitate the inference process. In our proposed method, $\sigma(s_{u,i})$ in Equation (\ref{eq:sentiment-effect}) captures the sentiment in the reviews. To validate the effectiveness of $\sigma(s_{u,i})$, we visualize the relationship between $\sigma(s_{u,i})$ and the sentiment polarity generated by TextBlob.
We provide the outcomes of our model on four datasets, as shown in Figure \ref{fig:senti-sigma-plot}. We plot the relationship between the average $\sigma(s_{u,i})$ and $s_u * s_i$,  which are both computed by TextBlob with reviews as input.  It is clear that there is a positive correlation between 
$\sigma(s_{u,i})$ and $s_u * s_i$ on all four datasets. On the Gourmet and Yelp datasets, there is a little descending slope in low sentiments. In all other parts, we see a strong positive correlation, which denotes $\sigma(s_{u,i})$ can capture sentiments.

\section{Conclusion}
In this work, we propose to address sentiment bias in RRSs by  counterfactual inference. We build a causal graph that treats the sentiment as a mediator variable and uses the Natural Direct Effect (NDE) to mitigate this bias. Extensive experiments are conducted, and we compare our method with the state-of-the-art methods. The results show that our method achieves a better performance on both rating prediction and sentiment bias mitigation.

\bibliography{aaai25}

\end{document}


\section{Technical Appendix}
\subsection{Technical Background Supplementary}
\begin{itemize}
    \item Reference state: As the goal of counterfactual inference in our method is to find out "What would the rating score be if RRSs were divested from sentiment bias?", reference state represents a baseline state where RRSs have not been affected by sentiment bias. 
    \item Null value: In counterfactual inference, null value refers to zero or invalid value when there is no specific intervention to determine the effect of the intervention.
\end{itemize}

During the debias phase, we cannot obtain the real value for null values like $u^*$ or $i^*$ due to the difficulty of getting the real value. Thus, we focus on the reference state $Y_{u^*,i^*}$ to solve this problem. We take the reference state into account by a hyper-parameter $\beta$ and search the optimal one during validation. Note that this $\beta$ is not equal to the reference state but $\beta = \gamma_{diccount} \cdot Y_{u^*,i^*}$.
This could be demonstrated in $Y_{u^*,i^*,S_{u,i}} = \beta \cdot \sigma(s_{u,i}) = Y_{u^*,i^*} \cdot \gamma_{diccount} \cdot \sigma(s_{u,i})$, where $\gamma$ is a factor to diminish the debiasing effect(the $\sigma(s_{u,i})$) since severe debiasing could dramatically decrease the recommendation performance. The specific values for $\beta$ is listed in the subsection Hyperparameter Setting below.



\subsection{Computing Infrastructure}

\begin{itemize}
    \item GPU models: NVIDIA TITAN X, NVIDIA TITAN RTX
    \item CPU models: Intel(R) Xeon(R) CPU E5-2697 v2 @ 2.70GHz
    \item Memory: 792GB
    \item OS: CentOS Linux 7 (Core)
    \item Software Libraries: Conda 4.10.3. Other dependent libraries are provided in a yaml file in the code appendix.
\end{itemize}

\subsection{Hyperparameter Setting}
In our experiment setting, we use the Adam optimizer, whose learning rate is 0.002. The weight decay parameter is $1 \times 10^{-6}$. We conduct a grid search for hyper-parameters $\alpha_u, \alpha_i$ and $\beta$. We search $\alpha$ in the range of $[1 \times 10^{-4}, 1.0]$ and find that the best performance is achieved when the $\alpha_u$ and $\alpha_i$ are 0.001. $\beta$ is searched in the set $\{0.01, 0.02, 0.03, 0.04, 0.05, 0.1, 0.2, 0.3, 0.4, 0.5, 0.6, 0.7\}$. We use a small $\beta$ because the MSE increases significantly when we use a larger $\beta$, indicating that there is a tradeoff between prediction accuracy and debiasing performance.

We conduct expeirments with preset random seeds that are in the set $\{670849,234725,300191,49002,237952\}$. These numbers are generated by Numpy random number generator with maximum $1\times 10^6$. We also conduct one experiment without setting the random seed. The results in Table \ref{Mean and variance} include both mean and variance of all our experiments.

\subsection{Wilcoxon Signed-rank Test}
We use the Wilcoxon signed-rank test to judge the significance of the improvement in our model's performance compared with CISD, because the differences are severely non-normally distributed. To test the null hypothesis that there is no performance difference between the models, we apply the Wilcoxon Signed-rank Test since the distribution of MSE differences between the two models is non-normally distributed. The test is conducted by using wilcoxon function from scipy.stats in Python. We set the alternative hypothesis as "greater" and obtain the approximate \textbf{\textit{p-value}} of 0.039, less than 0.05. Thus we reject the null hypothesis, indicating the significance of our proposed method. The details is shown in Table \ref{Wilcoxon Signed-rank}.

\begin{table}[ht]
\small
\centering
\begin{tabular}{cccccc}
\toprule
\textbf{Model}  & \textbf{CISD} & \textbf{Ours} & \textbf{Difference} & \textbf{Signed-Rank}\\ 
\midrule
 Gourmet           &0.9641   & 0.9478      &0.0163  &  2     &   \\ 
\midrule
 Video         &1.0830   &  1.0305     &0.0525   &   5    &  \\ 
\midrule
 Kindle      & 0.6104  & 0.5702      & 0.0402 &     3  &    \\ 
\midrule
 Elec        & 1.2253  &  1.2373     & -0.0120 &   -1   &   \\ 
\midrule
 Yelp         & 1.4473  & 1.3981     & 0.0492&    4   &   \\ 
\bottomrule
\end{tabular}
\caption{Wilcoxon Signed-rank Test on model performance.}
\label{Wilcoxon Signed-rank}
\end{table}

\begin{table*}[h!]
\centering
\begin{tabular}{c|c|c|c|c}
\toprule
\textbf{Datasets} & \textbf{Model}   & \textbf{BU} & \textbf{BI} & \textbf{MSE}    \\ 
\midrule
\midrule
\multirow{3}{*}{Gourmet}         & NARRE           & 1.2759  & 0.8067      & 0.9669   \\ 
& De-bias   & 1.2344  & 0.7655      & 0.9652   \\ 
& Ours            & $\textbf{1.1892} \pm 0.0140$  & $1.0169 \pm 0.0109$      & $\textbf{0.9478} \pm 0.0007$   \\ 
\midrule

\multirow{3}{*}{Video Games}         & NARRE      & 2.2774  & 2.1054 & 1.5260   \\ 
& De-bias    & 1.9752  & 1.4594      & 1.4388   \\ 
& Ours            & $\textbf{1.2010} \pm 0.0187$  & $\textbf{0.6221}\pm 0.0203$      & $\textbf{1.0305} \pm 0.0036$  \\ 
\midrule

\multirow{3}{*}{Kindle}         & NARRE           & 1.0024  & 0.7044  & 0.6612   \\ 
& De-bias    & 0.9247  & 0.6469    & 0.6244   \\ 
& Ours            & $\textbf{0.8508}\pm 0.0063$  & $\textbf{0.4701}\pm 0.0044$    & $\textbf{0.5702}\pm 0.0007$   \\ 
\midrule

\multirow{3}{*}{Electronics}         & NARRE      & 1.4132  & 1.1890  & 1.2588   \\ 
& De-bias    & 1.3522  & 1.1352      & 1.2394   \\ 
& Ours            & $\textbf{1.3043}\pm 0.0286$  & $\textbf{1.0812}\pm 0.0198$      & $\textbf{1.2373} \pm 0.0016$   \\ 
\midrule

\multirow{3}{*}{Yelp}         & NARRE           & 2.1043   & 1.2952   & 1.5119   \\ 
& De-bias   & 1.7749  & 1.0028      & 1.4535   \\ 
& Ours            & $\textbf{1.2844}\pm 0.0545$  & $\textbf{0.9343}\pm 0.0618$     & $\textbf{1.3981}\pm 0.0093$  \\ 

\bottomrule
\end{tabular}
\caption{BU, BI and MSE for RSS models on Amazon and Yelp Datasets.}
\label{Mean and variance}
\end{table*}

\subsection{Larger version of figures in Experiment}
\begin{figure*}[h]
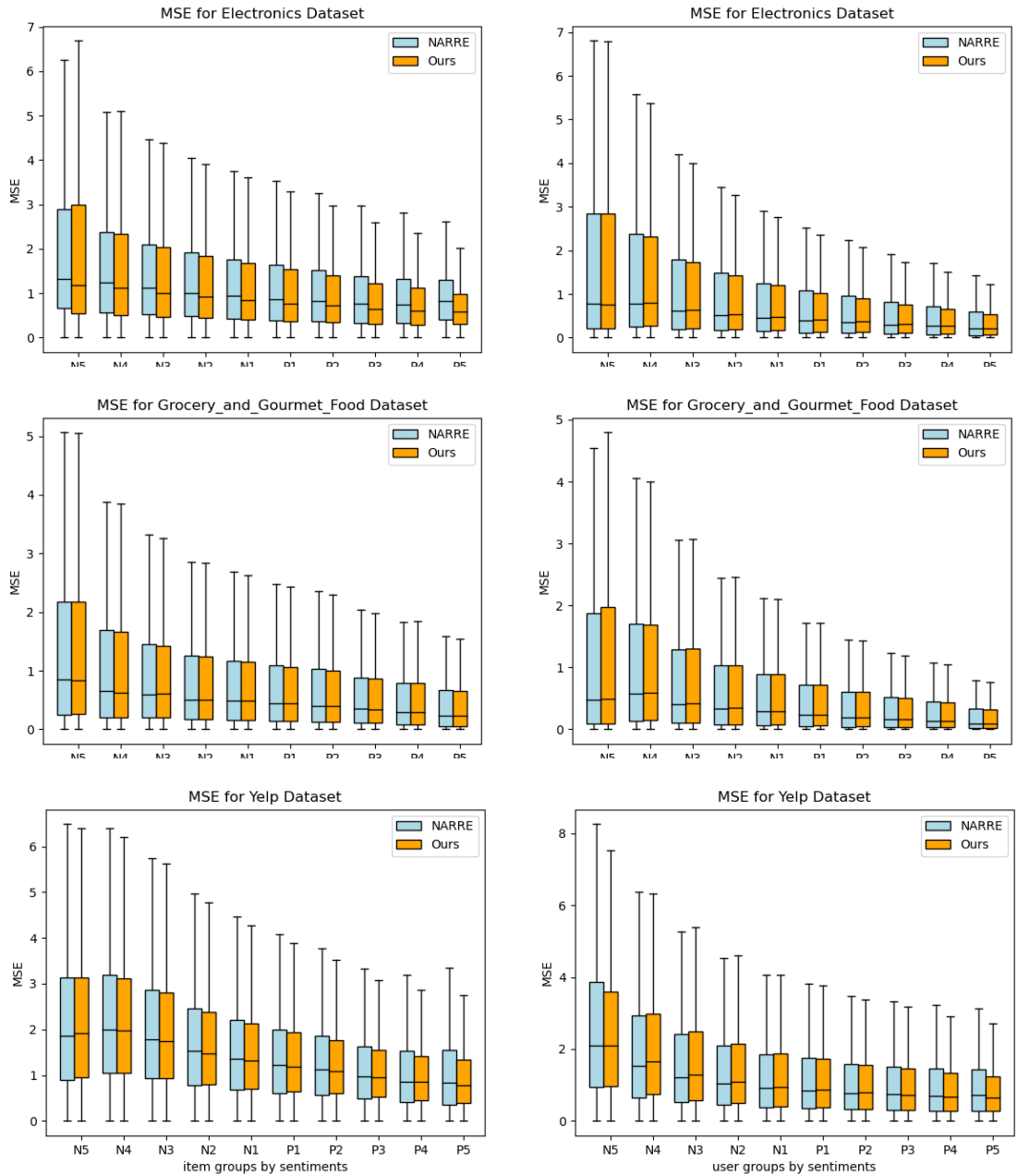

\centering
\includegraphics[scale = 0.5]{figure/boxplots/ci-multiple-boxplot-item-Electronics-NARRE_CI.png}\hspace*{-6mm}
\includegraphics[scale = 0.5]{figure/boxplots/ci-multiple-boxplot-user-Electronics-NARRE_CI.png}
\vspace*{-6mm}
\smallskip

\includegraphics[scale = 0.5]{figure/boxplots/ci-multiple-boxplot-item-Grocery_and_Gourmet_Food-NARRE_CI.png}\hspace*{-6mm}
\includegraphics[scale = 0.5]{figure/boxplots/ci-multiple-boxplot-user-Grocery_and_Gourmet_Food-NARRE_CI.png}
\vspace*{-6mm}
\smallskip

\includegraphics[scale = 0.5]{figure/boxplots/ci-multiple-boxplot-item-Yelp-NARRE_CI.png}\hspace*{-6mm}
\includegraphics[scale = 0.5]{figure/boxplots/ci-multiple-boxplot-user-Yelp-NARRE_CI.png}

\caption{Boxplots on mean MSE before and after debias(RQ3)}
\label{fig1}
\end{figure*}

\begin{figure*}[h]
\centering
\includegraphics[scale = 0.5]{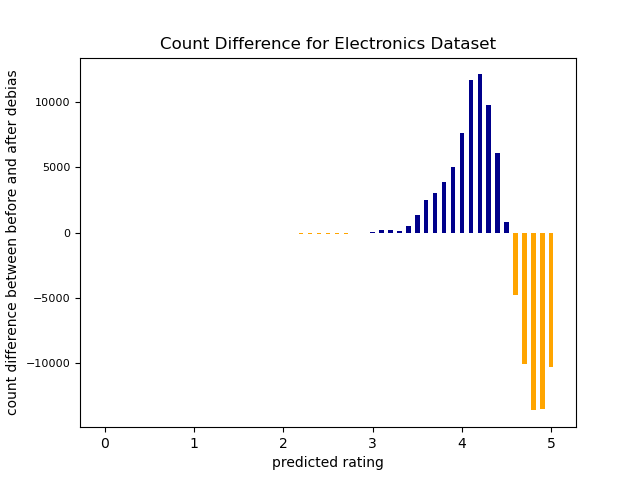}\hspace*{-6mm}
\includegraphics[scale = 0.5]{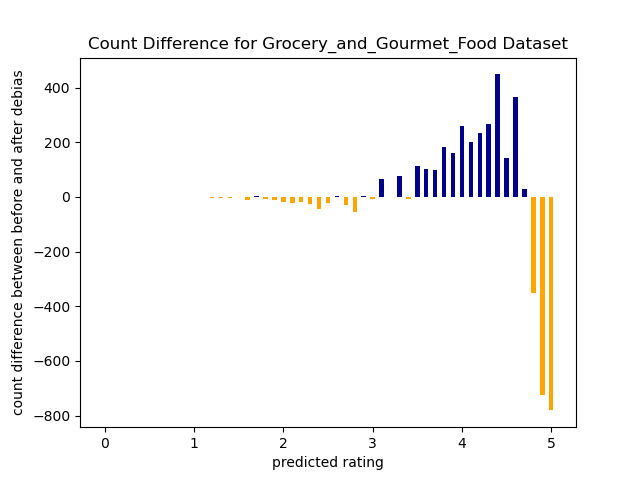}

\smallskip
\vspace*{-4.5mm}

\includegraphics[scale = 0.5]{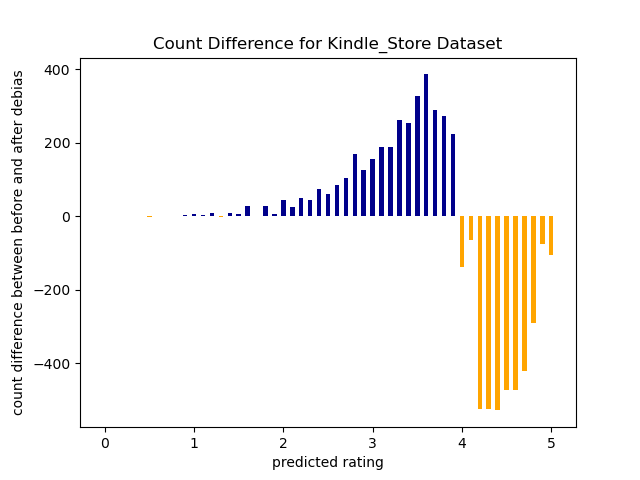}\hspace*{-6mm}
\includegraphics[scale = 0.5]{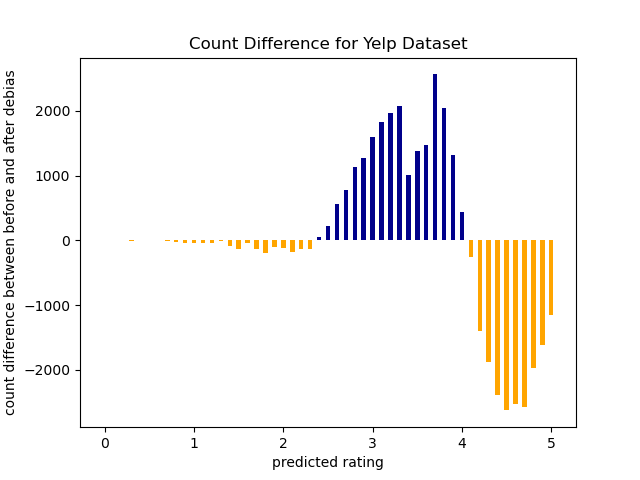}

\caption{Difference between before and after debias in rating distribution shift(RQ3).}
\label{fig2}
\end{figure*}

\begin{figure*}[h!]
\centering

\includegraphics[scale = 0.5]{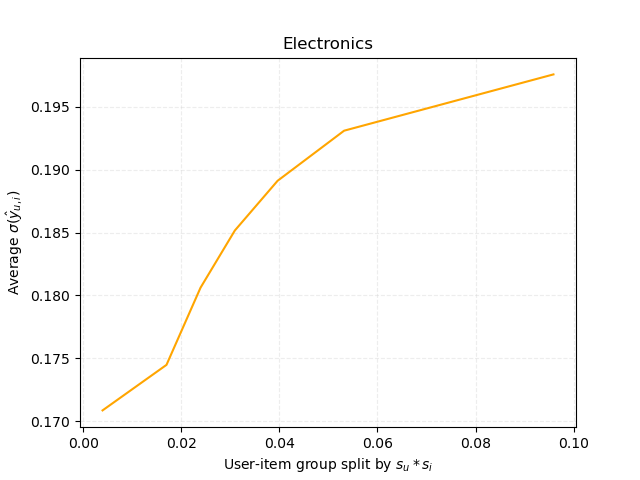}\hspace*{-6mm}
\includegraphics[scale = 0.5]{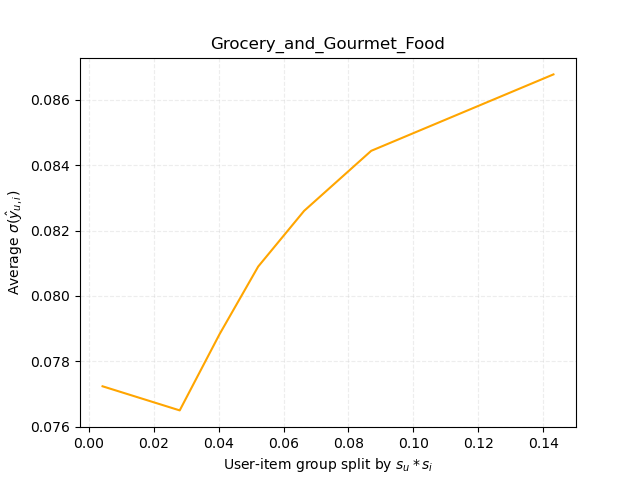}

\smallskip
\vspace*{-4.5mm}

\includegraphics[scale = 0.5]{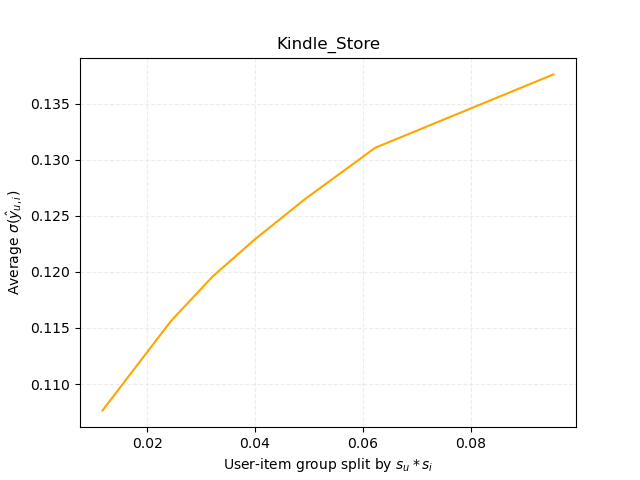}\hspace*{-6mm}
\includegraphics[scale = 0.5]{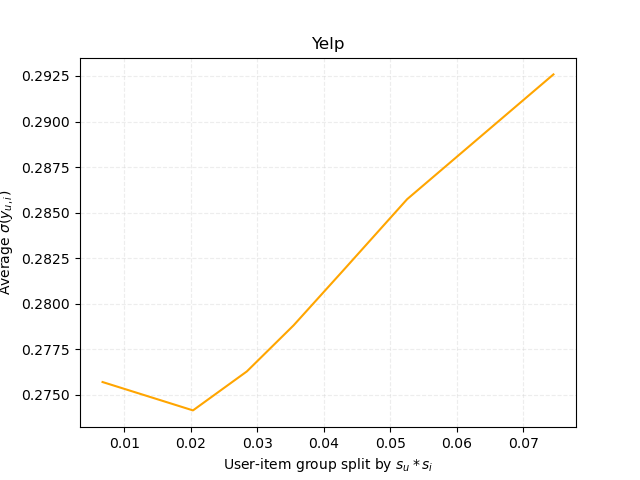}

\caption{The relationship between the predicted $\sigma(\hat{y}_{u,i})$ and related sentiment $s_u \cdot s_i$(RQ4).}
\label{fig3}
\end{figure*}

In order to present our experimental results more clearly with more details compared to the ones in the main body of our submitted manuscript,  we provide a larger version of figures obtained from the experiments, as shown in Figure \ref{fig1},\ref{fig2},\ref{fig3}.

\section{Reproducibility Checklist}
This paper:
\begin{itemize}
\item Question: Includes a conceptual outline and/or pseudocode description of AI methods introduced. (yes/partial/no/NA)
Answer: \textbf{YES}
\item Question: Clearly delineates statements that are opinions, hypothesis, and speculation from objective facts and results. (yes/no)
Answer: \textbf{YES}
\item Question: Provides well marked pedagogical references for less-familiare readers to gain background necessary to replicate the paper. (yes/no)
Answer: \textbf{YES}

\end{itemize}

\subsection{1. Contributions}

Question: Does this paper make theoretical contributions? (yes/no)
Answer: \textbf{YES}

\noindent Question: All assumptions and restrictions are stated clearly and formally. (yes/partial/no)
Answer:\textbf{YES}

\noindent Question: All novel claims are stated formally (e.g., in theorem statements). (yes/partial/no)
Answer: \textbf{YES}

\noindent Question: Proofs of all novel claims are included. (yes/partial/no)
Answer: \textbf{YES}

\noindent Question: Proof sketches or intuitions are given for complex and/or novel results. (yes/partial/no)
Answer: \textbf{YES}

\noindent Question: Appropriate citations to theoretical tools used are given. (yes/partial/no)
Answer: \textbf{YES}

\noindent Question: All theoretical claims are demonstrated empirically to hold. (yes/partial/no/NA)
Answer: \textbf{YES}

\noindent Question: All experimental code used to eliminate or disprove claims is included. (yes/no/NA)
Answer: \textbf{NA}

\subsection{2. Datasets}
\noindent Question: Does this paper rely on one or more datasets? (yes/no)
Answer: \textbf{YES}

\noindent Question: A motivation is given for why the experiments are conducted on the selected datasets (yes/partial/no/NA)
Answer: \textbf{YES}

\noindent Question: All novel datasets introduced in this paper are included in a data appendix. (yes/partial/no/NA)
Answer: \textbf{NA}

\noindent Question: All novel datasets introduced in this paper will be made publicly available upon publication of the paper with a license that allows free usage for research purposes. (yes/partial/no/NA)
Answer:\textbf{NA}

\noindent Question: All datasets drawn from the existing literature (potentially including authors’ own previously published work) are accompanied by appropriate citations. (yes/no/NA)
Answer: \textbf{YES}

\noindent Question: All datasets drawn from the existing literature (potentially including authors’ own previously published work) are publicly available. (yes/partial/no/NA)
Answer: \textbf{YES}

\noindent Question: All datasets that are not publicly available are described in detail, with explanation why publicly available alternatives are not scientifically satisficing. (yes/partial/no/NA)
Answer: \textbf{NA}

\subsection{2. Experiments}
\noindent Question: Does this paper include computational experiments? (yes/no)
Answer: \textbf{YES}

\noindent Question: Any code required for pre-processing data is included in the appendix. (yes/partial/no).
Answer: \textbf{YES}

\noindent Question: All source code required for conducting and analyzing the experiments is included in a code appendix. (yes/partial/no)
Answer: \textbf{YES}

\noindent Question: All source code required for conducting and analyzing the experiments will be made publicly available upon publication of the paper with a license that allows free usage for research purposes. (yes/partial/no)

\noindent Answer: \textbf{YES}

\noindent Question: All source code implementing new methods have comments detailing the implementation, with references to the paper where each step comes from (yes/partial/no)
\noindent Answer: \textbf{Yes}

\noindent Question: If an algorithm depends on randomness, then the method used for setting seeds is described in a way sufficient to allow replication of results. (yes/partial/no/NA)
\noindent Answer: \textbf{YES}

\noindent Question: This paper specifies the computing infrastructure used for running experiments (hardware and software), including GPU/CPU models; amount of memory; operating system; names and versions of relevant software libraries and frameworks. (yes/partial/no)
Answer: \textbf{YES}

\noindent Question: This paper formally describes evaluation metrics used and explains the motivation for choosing these metrics. (yes/partial/no)
\noindent Answer: \textbf{YES}

\noindent Question: This paper states the number of algorithm runs used to compute each reported result. (yes/no) Answer: \textbf{YES}

\noindent Question: Analysis of experiments goes beyond single-dimensional summaries of performance (e.g., average; median) to include measures of variation, confidence, or other distributional information. (yes/no)
Answer: \textbf{YES}

\noindent Question: The significance of any improvement or decrease in performance is judged using appropriate statistical tests (e.g., Wilcoxon signed-rank). (yes/partial/no)
Answer: \textbf{YES}

\noindent Question: This paper lists all final (hyper-)parameters used for each model/algorithm in the paper’s experiments. (yes/partial/no/NA)
Answer: \textbf{YES}

\noindent Question: This paper states the number and range of values tried per (hyper-) parameter during development of the paper, along with the criterion used for selecting the final parameter setting. (yes/partial/no/NA) Answer: \textbf{YES}

\newpage
\appendix